\def\ts     {\thinspace}
\def\kms    {\ifmmode{{\rm \ts km\ts s}^{-1}}\else{\ts km\ts s$^{-1}$}\fi}
\def\msol   {\ifmmode{{\rm M}_{\odot}}\else{M$_{\odot}$}\fi}
\def\lsol   {\ifmmode{{\rm L}_{\odot}}\else{L$_{\odot}$}\fi}
\def\zsol   {\ifmmode{{\rm Z}_{\odot}}\else{Z$_{\odot}$}\fi}
\def\etal   {{\rm et\ts al.}}
\def\aco    {\ifmmode{{\rm CO}(J\!=\!1\! \to \!0)}\else{{\rm CO}($J$=1$\to$0)}\fi}
\def\bco    {\ifmmode{{\rm CO}(J\!=\!2\! \to \!1)}\else{{\rm CO}($J$=2$\to$1)}\fi}
\def\cco    {\ifmmode{{\rm CO}(J\!=\!3\! \to \!2)}\else{{\rm CO}($J$=3$\to$2)}\fi}
\def\dco    {\ifmmode{{\rm CO}(J\!=\!4\! \to \!3)}\else{{\rm CO}($J$=4$\to$3)}\fi}
\def\eco    {\ifmmode{{\rm CO}(J\!=\!5\! \to \!4)}\else{{\rm CO}($J$=5$\to$4)}\fi}
\def\fco    {\ifmmode{{\rm CO}(J\!=\!6\! \to \!5)}\else{{\rm CO}($J$=6$\to$5)}\fi}
\def\gco    {\ifmmode{{\rm CO}(J\!=\!7\! \to \!6)}\else{{\rm CO}($J$=7$\to$6)}\fi}
\def\draco  {\ifmmode{^{13}{\rm CO}(J\!=\!1\! \to \!0)}\else{$^{13}${\rm CO}($J$=1$\to$0)}\fi}
\def\drbco  {\ifmmode{^{13}{\rm CO}(J\!=\!2\! \to \!1)}\else{$^{13}${\rm CO}($J$=2$\to$1)}\fi}
\def\drcco  {\ifmmode{^{13}{\rm CO}(J\!=\!3\! \to \!2)}\else{$^{13}${\rm CO}($J$=3$\to$2)}\fi}
\def\zwco   {\ifmmode{{\rm CO}}\else{CO}\fi}
\def\drco   {\ifmmode{^{13}{\rm CO}}\else{$^{13}$CO}\fi}
\def\ci     {\ifmmode{{\rm C}{\rm \small I}}\else{C\ts {\scriptsize I}}\fi}
\def\hi     {\ifmmode{{\rm H}{\rm \small I}}\else{H\ts {\scriptsize I}}\fi}
\def\hh     {\ifmmode{{\rm H}_2}\else{H$_2$}\fi}
\def\cone {\ifmmode{{\rm C}{\rm \small I}(^3\!P_1\!\to^3\!P_0)}
     \else{C\ts {\scriptsize I}{\small$(^3\!P_1\!\to^3\!\!\!P_0)$}}\fi}
\def\ctwo {\ifmmode{{\rm C}{\rm \small I}(^3\!P_2\!\to^3\!P_1)}
     \else{C\ts {\scriptsize I}{\small$(^3\!P_2\!\to^3\!\!\!P_1)$}}\fi}
\def\cij {\ifmmode{{\rm C}{\rm \small I}\,(^3P_i\to^3P_j)}\else{C\ts {\scriptsize I}\,{\small$(^3P_i\to^3P_j)$}}\fi}
\def\cii    {\ifmmode{{\rm C}{\rm \small II}}\else{C\ts {\scriptsize II}}\fi}
\def\tex {\ifmmode{{T}_{\rm ex}}\else{$T_{\rm ex}$}\fi}
\def\tmb {\ifmmode{{T}_{\rm mb}}\else{$T_{\rm mb}$}\fi}
\def\tkin {\ifmmode{{T}_{\rm kin}}\else{$T_{\rm kin}$}\fi}
\def\microns {\ifmmode{\mu{\rm m}}\else{$\mu$m}\fi}
\def\nhh   {\ifmmode{n({\rm H}_2)}\else{$n$(H$_2$)}\fi}
\def\gradv {\ifmmode{{\rm dv/dr}}\else{dv/dr}\fi}
\documentclass[]{aa}
\usepackage{graphicx}
\begin{document}
   \title{The Spectral Energy Distribution of CO lines in M\,82}

   \author{A. Wei\ss
          \inst{1}
          \and
          F. Walter
          \inst{2}
          \and 
          N.Z. Scoville
          \inst{3}
          }

   \offprints{A. Wei\ss}

   \institute{IRAM, Avenida Divina Pastora 7, 18012 Granada, Spain\\
              \email{aweiss@iram.es}
         \and
             MPIA, K\"onigstuhl 17, 69117 Heidelberg, Germany\\
             \email{walter@mpia.mpg.de}
         \and
             Caltech, Pasadena, CA, USA\\
             \email{nzs@astro.caltech.edu}
             }

   \date{April 4, 2005}

   \abstract{We present an analysis of the excitation conditions of the molecular
gas in the streamers and the outflow of M\,82 based on
observations obtained at the IRAM 30\,m telescope. Our analysis
of $J$=1$\to$0 and $J$=2$\to$1 transitions of \zwco\ and \drco\ and the
\cco\ line in 13 regions outside the central starburst disk shows that
the gas density within the streamer/outflow system is about an order
of magnitude lower ($log(n_{\hh}) \approx 3.0\,{\rm cm}^{-3}$) than in
the central molecular disk. We have used an LVG model and data from
the literature to constrain the flux density in each CO transition (the `CO
line SED') arising from the streamer/outflow system and the central
starburst disk itself. 
Globally, we find that the CO flux density up to the $J$=3$\to$2
line is dominated by the diffuse outer regions while lines above the
$J$=5$\to$4 transition are almost exclusively emitted by the central starburst
disk. We compare the CO line SED of M\,82 to CO observations
of galaxies at high redshift and suggest that small high--$J$/low--$J$ CO flux
density ratios (observed in some of these sources) are not necessarily 
caused by a different excitation of the central molecular gas
concentration, but may result from an additional, more extended and
diffuse gas reservoir around these systems, reminiscent of the situation in M\,82.

   }

   \maketitle
%

\section{Introduction}
Studying the physical properties of the molecular gas 
in galaxies is of fundamental importance for
understanding the processes leading to star formation at low and
high redshifts. Due to its proximity and its strong 
emission in molecular lines, the nearby starburst galaxy M\,82 is 
one of the best studied starburst environments. The central concentration of
molecular gas, which feeds the strong star
formation activity, has been the subject of many studies
addressing both the excitation conditions and the distribution of the
molecular gas (e.g., Young \& Scoville \cite{young84}, Wild \etal\ \cite{wild92}, Wei\ss\ \etal\
\cite{weiss99}, Mao \etal\ \cite{mao00}, Wei\ss\ \etal\ \cite{weiss01}, Ward \etal\
\cite{ward03}). Some of these studies already have provided evidence for
extended CO emission around M\,82's center (e.g. Stark \& Carlson
\cite{stark84}, Seaquist \& Clark \cite{seaquist01}, Taylor, Walter,
\& Yun \cite{taylor01}).  A high--resolution, wide--field CO study
revealed that large amounts of molecular gas are present in molecular
streamers and the outflow around M\,82's central disk (Walter, Wei\ss\
\& Scoville \cite{walter02}).

Recent studies of CO line and dust continuum emission  
in distant quasar host galaxies ($z>2$) have used M\,82's central
starburst disk as a nearby template (e.g. Yun \etal\ \cite{yun00}, Cox \etal\
\cite{cox02}, Carilli \etal\ \cite{carilli02a}, Carilli \etal\
\cite{carilli02b}, Wei\ss\ \etal\ \cite{weiss03}). At cosmological distances,
however, only integrated values can be derived as usually no
information on the spatial distribution of the molecular gas exists.
Therefore CO line ratios can only provide information on the galaxies average,
global excitation conditions.  Consequently a comparison of such line
ratios to values determined {\em only for the center of M\,82} are
incomplete and biased as they neglect the contribution of a possible
extended molecular gas component which could potentially have
significantly different excitation conditions. This is the main
motivation for the present study in which we present an analysis of
the excitation conditions of the extended molecular gas surrounding
M\,82's central starburst disk.

\section{Observations and data reduction}

\begin{figure*}
   \centering
   \includegraphics[width=14.0cm]{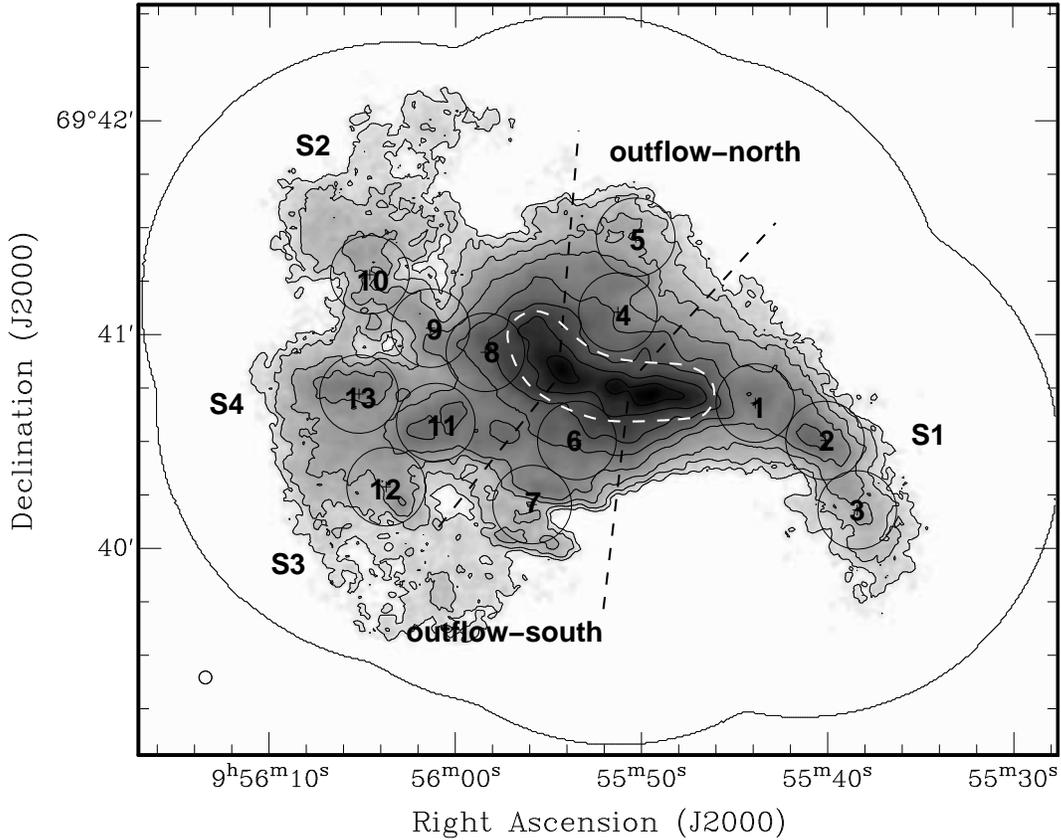}
   \caption{Logarithmic representation of the integrated \aco\ 
  map of the zero--spacing--corrected OVRO mosaic (Walter, Wei\ss\, \&
 Scoville 2002). Circles indicate the observed positions for which we analyzed the CO line brightness
  temperature ratios. The size of the circles corresponds to 22$''$,
  i.e. the spatial resolution of the 30\,m telescope at 3\,mm. The white
   dashed contour shows our division between the central molecular
   disk and the streamer/outflow system.}
   \label{positions}
   \end{figure*}

\begin{figure*}
   \centering
   \includegraphics[width=18.0cm]{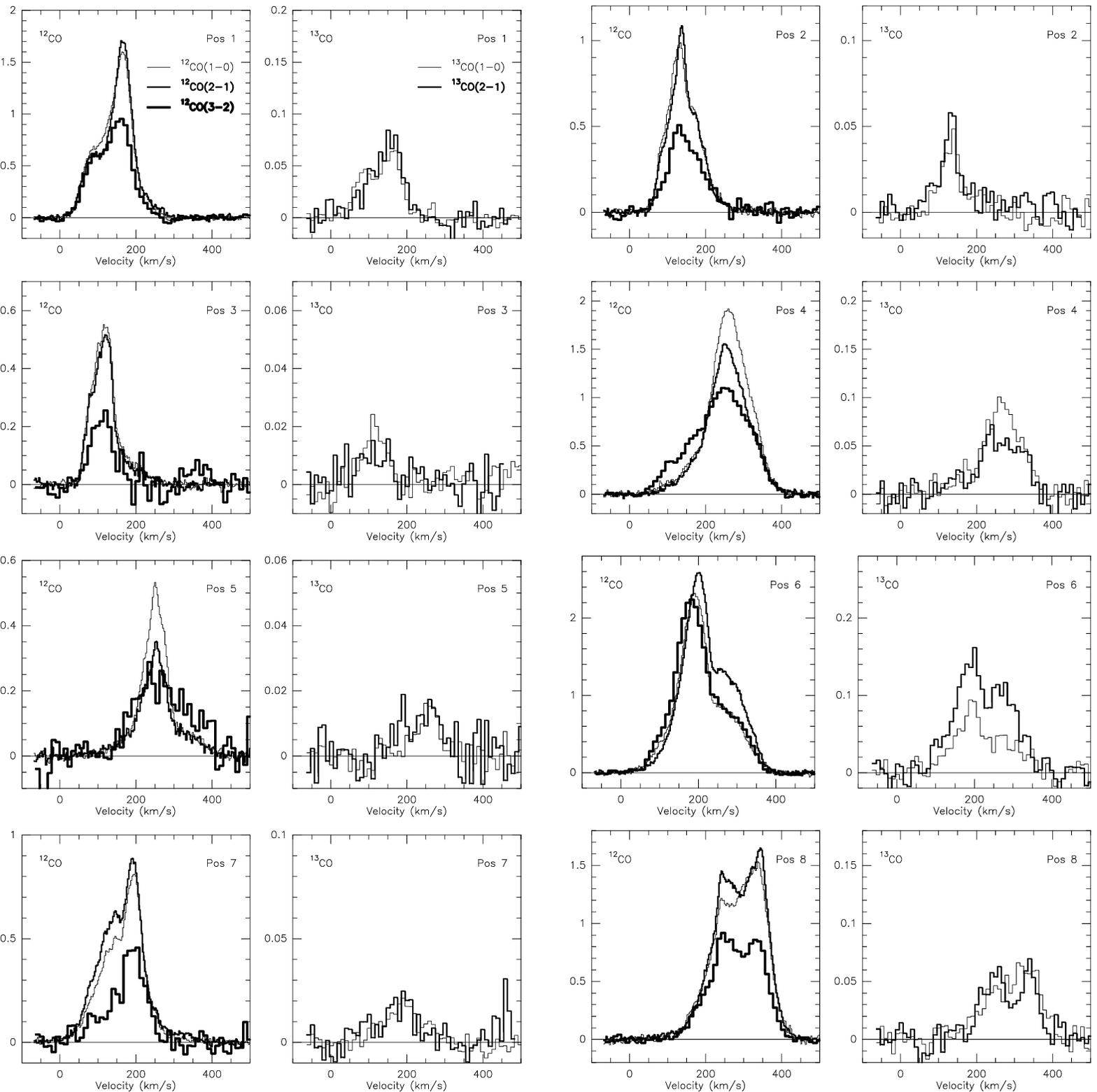}
   \caption{Spectra of the \aco, \bco, and \cco\ (left),
   and the \draco\ and \drbco\ (right) transitions at 22$''$ resolution at
   individual positions.  Line brightness temperatures are given in
   Kelvin on a $T_{\rm mb}$ scale. The \cco\
   spectra have been calculated from the data in Dumke \etal\ \cite{dumke01}.}
   \label{spectra1}
   \end{figure*}

Using the IRAM 30\,m telescope we observed the \aco, \bco, \draco\ and \drbco\ 
transitions towards 13 positions covering the molecular outflow and
the molecular streamers surrounding M\,82's starburst toroid (see Fig.~\ref{positions}). 
Spectra were obtained using
the wobbler switch technique with a wobbler throw of 240$''$ and a 
wobbler frequency of 0.5\,Hz. The receiver alignment was determined 
using pointing scans on Saturn and was found to be better 
than 2$''$. Pointing was checked every ~2h on the nearby quasars 
0954+658 and 0836+710. We estimate the pointing accuracy to be better 
than 3$''$. Calibration was checked once per observing run on
W3OH and the south-western molecular lobe in M\,82.\\
System temperatures were on average $T_{sys}$\,$\approx$\,180\,K and
330\,K at 115 and 230\,GHz 
respectively. The data were recorded using the Vespa auto correlator
with 512 MHz bandwidth and 1.25 MHz resolution at 3\,mm and two 1\,MHz filter banks (512 
channels each) at 1\,mm. The velocity coverage at 115\,GHz is 1330 \kms\ 
with 3.4 \kms\ resolution. The corresponding values at 230\,GHz are 660 \kms\ and 1.3 \kms. 
The telescope's resolution 
at 230 and 115\,GHz is 11$''$ and 22$''$ respectively. To enable
comparison between the brightness temperatures at 230 and 115\,GHz
we observed a cross consisting of five pointings at each position
(separation 6$''$). Depending on the CO line strength
the integration time at each central pointing was between 12 and
54\,min. All other pointings within the cross were observed for 6 or 12\,min.\\
For data processing we used the CLASS software package. Only linear
baseline were removed from each spectrum. For the \aco\, and the \draco\
lines only spectra from the central pointing were coadded. For the
reduction of the \bco\, and the \drbco\ data, spectra at each pointing
in the cross were coadded. From the resulting 5 spectra we generated
the final spectrum at 22$''$ resolution by a weighted addition with 
a weight of 1 for the central and 0.71 for the other four spectra
(Johannsson \etal\ \cite{johansson94}).   
The coadded spectra were 
converted to the main beam brightness temperature scale 
($T_{\rm mb} = F_{\rm eff}/B_{\rm eff}\,T_{A}^{*}$) using  $F_{\rm eff}
= 0.95$, $F_{\rm eff} = 0.91$ and $B_{\rm eff} = 0.71$, $B_{\rm eff} = 0.49$ at 115 and 230\,GHz
respectively (IRAM newsletter 55, 2003). Finally, \aco\, and \bco\, spectra were smoothed
to 3 \kms\ resolution. The final resolution of the \draco\, and \drbco\,
spectra is 10 \kms. In addition to theses observations we used the
\cco\ data from Dumke \etal\ (\cite{dumke01}). To allow a
comparison to our data we smoothed the \cco\ data cube to 22$''$
resolution before extracting spectra at each position covered by
our observations. Individual spectra at 22$''$ resolution are shown
in Fig.~\ref{spectra1}. Table \ref{lines} summarizes
the line parameters at each position.
\begin{figure*}
   \centering
   \includegraphics[width=18.0cm]{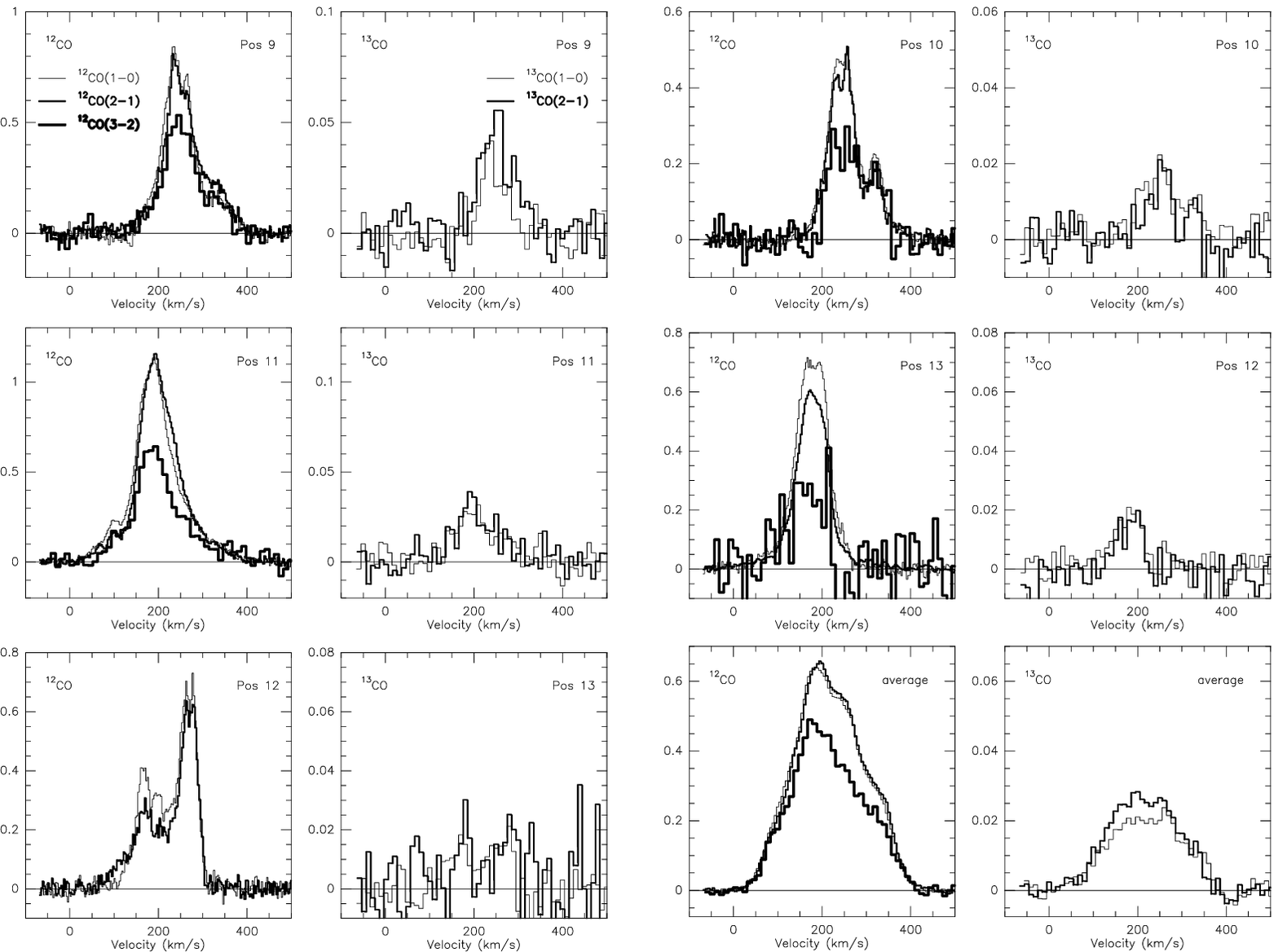}
   {\bf Fig.2.} continued. Note that the \cco\ spectrum at
   position 11 has a offset of $-5'',0''$ with respect to the other
   lines due the limited extend of the \cco\ cube.
   The spectra at the bottom right panels are average spectra from all positions.
   \end{figure*}
\begin{table*}
\caption[]{CO line parameters at 22$''$ resolution.}
\label{lines}
\begin{tabular}{r c c c c c c c c c c c}
\hline
\noalign{\smallskip}
& T$^{12}_{(1-0)}$ &T$^{12}_{(2-1)}$ &$^{a}$T$^{12}_{(3-2)}$ &T$^{13}_{(1-0)}$ &T$^{13}_{(2-1)}$ 
& W$^{12}_{(1-0)}$ & W$^{12}_{(2-1)}$ & $^{a}$W$^{12}_{(3-2)}$ &W$^{13}_{(1-0)}$&W$^{13}_{(2-1)}$
&v$_{peak}$\\ 
&[mK]&[mK]&[mK]&[mK]&[mK]&[K\kms]&[K\kms]&[K\kms]&[K\kms]&[K\kms]&[\kms]\\\hline
 1& 1590& 1690& 930& 60& 70& 155& 153& 107& 6.7& 7.0& 165\\
 2&  980& 1060& 450& 40& 50&  90&  87&  52& 3.5& 3.3& 130\\
 3&  540&  510& 250& 22& 11&  43&  38&  22& 1.7& 1.4& 120\\
 4& 1910& 1550&1090& 95& 60& 245& 196& $^{b}$197& 12& 8.1& 260\\
 5&  520&  340& 200& 15& 10&  44&  33&  $^{b}$35& 1.5& 1.4& 250\\
 6& 2310& 2580&2200& 90&140& 280& 335& 298& 12& 23& 190\\
 7&  810&  880& 440& 20& 19&  90&  104&  45&  2.3& 2.0& 195\\
 8& 1510& 1640& 890& 60& 57& 237& 253& 150& 10& 8.4& 330\\ 
 9&  840&  780& 500& 38& 52&  74&  74&  54&  3.1&  4.3& 235\\
10&  480&  480& 250& 19& 17&  47&  43&  28&  1.8&  1.3& 250\\
11& 1130& 1150& $^{c}$620& 28& 36& 128& 126&  $^{c}$80& 3.6& 3.5& 190\\
12&  700&  600& 260& 20& 20&  70&  56&  25& 1.9& 1.9& 180\\
13&  690&  610& -- & 17&$<$\,20&65&56&  -- &2.2&$<2.4$& 270\\ \hline
avg.&655 & 645 &480 &20 &26 &98 &100 &74 &3.6 &4.8 & \\
\end{tabular}

\begin{list}{}{}
\item[$^{\mathrm{a}}$] \cco\ data from Dumke \etal\ \cite{dumke01}
\item[$^{\mathrm{b}}$] \cco\ line profile wider than \aco\ and \bco\
line (see discussion in Sec.\,\ref{ratios-results})
\item[$^{\mathrm{c}}$] \cco\ spectrum has a offset of $-5'',0''$ from Pos.\,11
\end{list}
\end{table*}

\section{Results}

\subsection{Line brightness temperature ratios \label{ratios-results}}

The \aco, \bco\ and \draco\ lines are detected towards all
positions. At position 13 (streamer S4) the \drbco\ line is not
detected (3$\sigma$ upper limit: $T_{\drbco}< 20$\,mK).
The \cco\ line is detected at all positions except positions 13 which
is outside the field observed by Dumke \etal\ (\cite{dumke01}). The
line profiles of all transitions at individual positions 
are in good agreement except for the northern outflow (Pos. 4\,\&\,5) where the wings
of the \cco\ line are broader than the corresponding profiles for the
\aco\ and \bco\ lines (see Fig.~\ref{spectra1}). 
We attribute these line profile differences to small pointing
offsets between the observations by Dumke \etal\ and our observations.

We have used our high-resolution
CO map (see Fig.~\ref{positions}) and the beam
patterns of the 30\,m telescope (Greve, Kramer \& Wild \cite{greve98}) 
to estimate the contribution from the error--beam in each spectrum.
Using the first two error-beam patterns we find that the contribution
is negligible at all positions ($<$5\%).

The integrated and peak brightness temperature ratios agree well
within the observational errors. In the following we use the
{\em integrated} brightness temperature ratios, since this provides a higher
signal to noise ratio for the $^{13}$CO data. All $R$ ratios used
throughout this paper are defined in Table \ref{lineratios}.
Only in the case of $R_{31}$ in the northern outflow (Pos. 4\,\&\,5) 
we used the {\em peak} brightness temperature ratio, as the 
integrated \cco\ intensities are too high due to the different
line profile (see above). Statistical
errors are negligible for all CO lines. From our W3OH scans we
estimate the systematic error of the $R_{10}$ ratio to be $\approx 15\%$. The different observing
methods at 3\,mm and 1\,mm lead to  $\approx 30\%$ systematical errors
for $R_{21}$ and $R^{13}_{21}$. Taking the independent uncertainties
of our \aco\ intensities and the \cco\ errors reported by Dumke \etal\ 
(\cite{dumke01}) lead to $\approx 30\%$ error for the
$R_{31}$ ratio. Line brightness temperature ratios and their errors, taking the
statistical error for the $^{13}$CO lines into account, are summarized
in Table\ \ref{lineratios}.

\begin{table*}
\caption[]{Selected integrated line brightness temperature ratios at 22$''$
resolution. Errors include 15\% systematic uncertainty for $R_{10}$,
30\% for $R_{21}$ and $R^{13}_{21}$ and 30\% for $R_{32}$. Statistical
errors have been included for the $^{13}$CO lines only.}
\label{lineratios}
\begin{tabular}{r c c c c c c c c l}
\hline
\noalign{\smallskip}
 & $\alpha$ &  $\delta$
 & $\frac{W({\rm CO}(2-1))}{W({\rm CO}(1-0))}$
 & $\frac{W({\rm CO}(3-2))}{W({\rm CO}(1-0))}$ 
 & $\frac{W(^{13}{\rm CO}(2-1))}{W(^{13}{\rm CO}(1-0))}$ 
 & $\frac{W({\rm CO}(1-0))}{W(^{13}{\rm CO}(1-0))}$ 
 & $\frac{W({\rm CO}(2-1))}{W(^{13}{\rm CO}(2-1))}$
 & $\frac{W({\rm CO}(2-1))}{W(^{13}{\rm CO}(1-0))}$
 & region\\
 & \multicolumn{2}{c}{J2000.0}& $R_{21}$ & $R_{31}$  & $R^{13}_{21}$&
 $R_{10}$& $R^{12/13}_{2}$& $R^{12/13}_{21}$&\\  
\noalign{\smallskip} \hline
1 & 09:55:43.9 & 69:40:41 & $1.0\pm0.3$ & $0.7\pm0.2$  &  $1.0\pm0.4$ &$23\pm5$ & $22\pm5$&$23\pm5$&S1\\
2 & 09:55:40.1 & 69:40:30 & $1.0\pm0.3$ & $0.6\pm0.1$  &  $0.9\pm0.4$ &$26\pm6$ & $26\pm4$&$25\pm5$&S1\\
3 & 09:55:38.4 & 69:40:11 & $0.9\pm0.3$ & $0.5\pm0.1$  &  $0.8\pm0.2$ &$25\pm6$ & $27\pm6$&$22\pm5$&S1\\
4 & 09:55:51.3 & 69:41:06 & $0.8\pm0.25$& $^{a}0.6\pm0.2$&$0.7\pm0.2$ &$20\pm4$ & $24\pm5$&$16\pm5$&O--N\\
5 & 09:55:50.3 & 69:41:27 & $0.7\pm0.15$&$^{a}0.4\pm0.15$&$0.9\pm0.4$ &$29\pm7$ & $24\pm12$&$22\pm5$&O--N\\
6 & 09:55:53.5 & 69:40:30 & $1.2\pm0.3$ & $1.1\pm0.2$  &  $1.5\pm0.6$ &$22\pm6$ & $19\pm4$&$28\pm6$&O--S\\
7 & 09:55:55.9 & 69:40:12 & $1.1\pm0.3$ & $0.5\pm0.2$  &  $0.9\pm0.4$ &$39\pm10$& $54\pm10$&$45\pm9$&O--S\\
8 & 09:55:58.4 & 69:40:55 & $1.0\pm0.3$ & $0.6\pm0.2$  &  $0.8\pm0.4$ &$24\pm6$ & $30\pm5$&$25\pm5$&S2\\
9 & 09:56:01.3 & 69:41:02 & $1.0\pm0.3$ & $0.7\pm0.2$  &  $1.4\pm0.5$ &$24\pm6$ & $17\pm3$&$24\pm5$&S2\\
10& 09:56:04.6 & 69:41:17 & $0.9\pm0.3$ & $0.6\pm0.2$  &  $0.7\pm0.4$ &$26\pm9$ & $33\pm5$&$24\pm5$&S2\\
11& 09:56:01.1 & 69:40:35 & $1.0\pm0.3$ & $^{b}0.6\pm0.15$ &  $1.0\pm0.5$ &$36\pm10$& $36\pm8$&$35\pm8$&S3\\
12& 09:56:03.7 & 69:40:17 & $0.8\pm0.25$& $0.4\pm0.1$  &  $1.0\pm0.4$ &$36\pm14$& $29\pm6$&$29\pm6$&S3\\
13& 09:56:05.2 & 69:40:43 & $0.9\pm0.25$&      ---     &  $<1.1$      &$30\pm10$& $>23   $&$25\pm5$&S4\\
\noalign{\smallskip} \hline				 							
  &            &          & $1.0\pm0.1$  & $0.8\pm0.2$ &  $1.3\pm0.2$
  &$27\pm7$ & $21\pm6$&$28\pm5$&$^{c}$ IWA\\
\end{tabular}
\begin{list}{}{}
\item[$^{\mathrm{a}}$] Peak brightness temperature ratio (see text). 
\item[$^{\mathrm{b}}$] \cco\ spectrum has a offset of $-5'',0''$ from Pos.\,11
\item[$^{\mathrm{c}}$] Intensity weighted average determined from the
  integrated line brightness temperature ratios of the averaged
  spectra (see Fig.\,\ref{spectra1} bottom right panels).
\end{list}
\end{table*}

\begin{figure*}
   \centering
   \includegraphics[width=14.0cm]{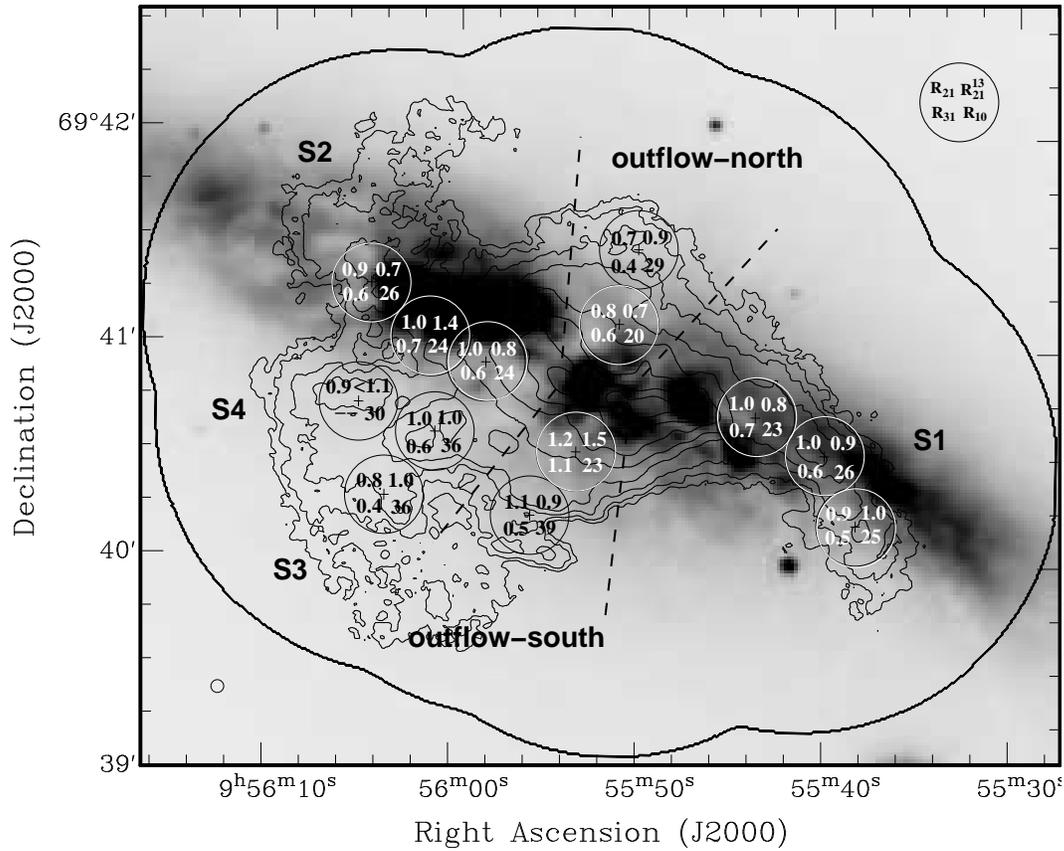}
   \caption{Spatial variations of the $R_{21}$ (upper left number),
   $R_{31}$ (lower left),
   $R^{12/13}_{21}$ (upper right) and $R_{10}$ (lower right) line brightness temperature
   ratios (see Table\ \ref{lineratios} for definitions). 
The greyscale is an optical V-band image of
   M\,82 (Cheng \etal\ \cite{cheng97}). The contours represent the
   distribution of the \aco\ emission (see Fig.\, \ref{positions}).}
   \label{ratio-var}
   \end{figure*}

Fig.\,\ref{ratio-var} shows the spatial variations of the
most relevant line brightness temperature ratios. Differences
between individual positions are remarkably small. The only clear trend
visible in the spatial distribution of the line ratios is a decreasing
$R_{31}$ with galactocentric distance. This trend is visible along
the streamer S1 and S3 as well as along the northern and southern
outflow. Along S2, however, $R_{31}$ is
constant. All other ratios do not show a clear trend with increasing
distance from the central starburst region. $R_{21}$ is around unity
for all positions covering the streamer system, slightly above unity in the
southern, but only about 0.8 in the northern outflow. The corresponding ratios
in $^{13}$CO, $R^{13}_{21}$, show a larger scatter and values range between 0.7 and 1.5. 
Judging from the ratio of the ground transitions of \zwco\ and
\drco, $R_{10}$, the analyzed positions can be subdivided into 
positions within and outside the optical disk. Within the disk,
along the molecular streamers S1 \& S2 and at the base of the northern and southern
outflow the ratio $R_{10}$ is in the range of $20-26$. Outside the optical disk, 
at the end the outflows and within the streamer system S3 and S4 the
ratio rises to value between $29-39$. While the line
ratios at the base of the southern outflow almost resemble those
in the molecular disk of M\,82 (see Sec. \ref{center}), the gas in the 
northern outflow appears to be much less excited. In fact,
$R_{21}$ and $R_{31}$ in the northern outflow are even
lower than in the molecular streamers. 

\subsection{Excitation Conditions \label{excitation}}

\subsubsection{Average Properties \label{average-excitation}}

To analyze the excitation conditions of the molecular gas surrounding
M\,82's starburst region we use a spherical, isothermal one
component large velocity gradient (LVG) model.  Given the small
variations of the line brightness temperature ratios with position 
we use in the following the line ratios determined from the average
spectra to investigate the global excitation away from the star
forming regions. Note that these ratios differ from the numerical average
of the line ratios at individual positions since they correspond to
the intensity weighted averages which would be observed if M\,82 was 
shifted to cosmological distances and the emission would remain unresolved.
Spatial variations of the excitation conditions
are discussed in section \ref{individual-excitation}. The averaged spectra are
shown at the bottom right panels of Fig. \ref{spectra1}, the corresponding
averaged integrated line ratios are given in the bottom row of Table \ref{lineratios}. 

We compare LVG predicted line ratios to the observations in a
LVG parameter space covering densities and kinetic temperatures of
$log(n_{\hh})= 2.0-5.0\,{\rm cm}^{-3}$ and $T_{\rm kin}=5-200$\,K. 
CO abundances per velocity gradient and $^{13}$CO abundances
relative to CO are in the range of [CO]/\gradv\ = $1\cdot10^{-6}
- 2\cdot10^{-4}$\,pc\,(\kms)$^{-1}$ and [CO]/[$^{13}$CO] = 30\,--\,100.\\ 
The observed \zwco\ ratios, $R_{21}$ and $R_{31}$, limit solutions in
the LVG parameter space to a density range between $log(n_{\hh})= 2.3-3.5\,{\rm cm}^{-3}$
where the lower density limit corresponds to the highest value of
[CO]/\gradv\ and the upper density limit to the lowest [CO]/\gradv. The measured high $R_{10}$ ratio restricts
solutions to physical conditions with moderate opacities of the \aco\ 
($\tau_{12} \approx 1-3$) and very low opacities in the \draco\ transition
($\tau_{13} \approx 0.01-0.05$) within this density range. As a 
consequence solutions are limited to small CO abundances per velocity gradient 
($\le 2\cdot10^{-5}$\,pc\,(\kms)$^{-1}$) and [\zwco]/[\drco] abundance
ratios $> 50$. This additional constraint on [CO]/\gradv\ yields an
allowed density range of $log(n_{\hh})= 2.7-3.5\,{\rm cm}^{-3}$ 
and small CO column densities per velocity interval ($<2\cdot10^{17}$\,cm$^{-2}$\,(\kms)$^{-1}$).
The kinetic temperature is not well constrained without assumptions on
the values of the [CO]/[$^{13}$CO] abundance ratio and the CO
abundance per velocity gradient. A lower limit on the kinetic
temperature of $T_{\rm kin}\ge30$\,K is given by the densest possible 
solution within the parameter space. 

\begin{figure*}
   \centering
   \includegraphics[width=18.0cm]{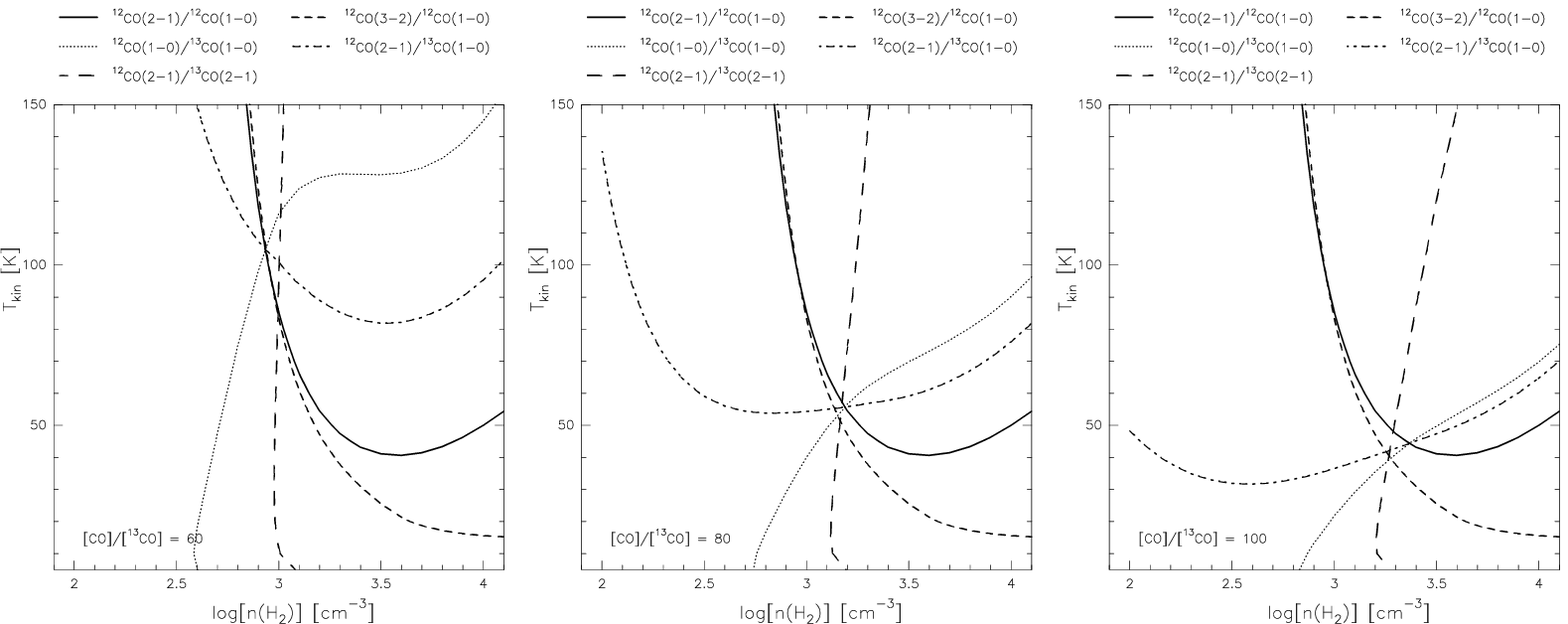}
   \caption{LVG line brightness temperature ratios as a function of
   the kinetic temperature and \hh\ density for a fractional $^{13}$CO
   abundance of 60, 80 and 100. The CO abundance per velocity gradient
 is $1\cdot10^{-5}$\,pc\,(\kms)$^{-1}$ for each graph. Displayed line
   ratios correspond to the averaged observed values listed in the
   bottom row of Table \ref{lineratios}.}
   \label{averaged-lvg1}
   \end{figure*}

\begin{figure*}
   \centering
   \includegraphics[width=18.0cm]{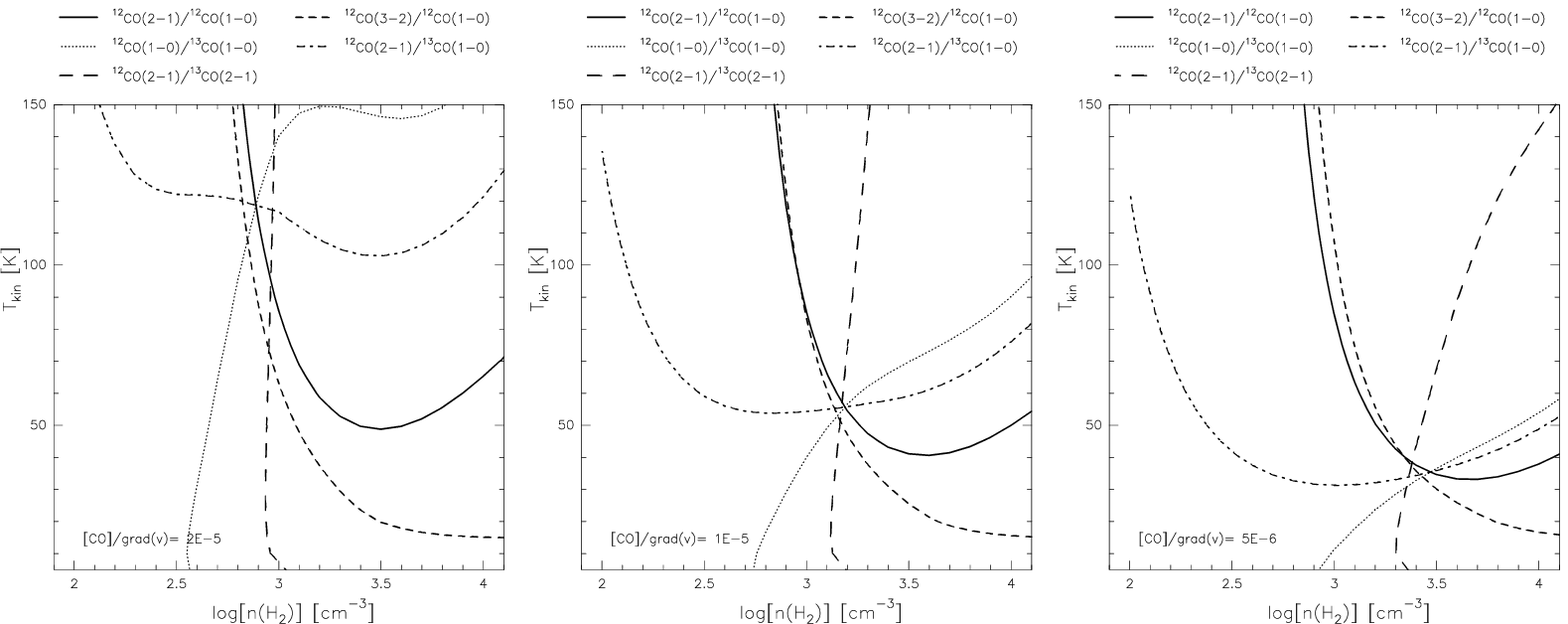}
   \caption{LVG line brightness temperature ratios as a function of
   the kinetic temperature and \hh\ density for a CO
   abundance per velocity gradient of $2\cdot10^{-5}$, 
   $1\cdot10^{-5}$ and $5\cdot10^{-6}$\,pc\,(\kms)$^{-1}$. The fractional
   $^{13}$CO abundance is 80 for each graph. Displayed line
   ratios correspond to the averaged observed values listed in the
   bottom row of Table \ref{lineratios}.}
   \label{averaged-lvg2}
   \end{figure*}

In Figs.\,\ref{averaged-lvg1} \& \ref{averaged-lvg2} we show LVG
solutions for the averaged line ratios as a function of abundance
and velocity gradient variations. The lines represent the
temperature--density combinations for which the LVG predicted line
ratios match the average observed line ratios. The figures show that 
lower \zwco\ abundances per velocity gradient and/or higher
fractional \drco\ abundances lead to cooler, denser solutions while
higher abundances per velocity gradients and/or lower
fractional \drco\ abundances lead to warmer and more diffuse solutions, respectively.
For a CO abundance per velocity gradient of
$1\cdot10^{-5}\,{\rm pc}\,(\kms)^{-1}$ and a [CO]/[$^{13}$CO] abundance ratio of 80
the averaged line intensity ratios are well reproduced by a kinetic
temperate of $T_{\rm kin} = 55$\,K with a \hh\ density of
$log(n_{\hh})= 3.2\,{\rm cm}^{-3}$. These values correspond to the
intersection of the lines shown in Fig.\,\ref{averaged-lvg1} 
\& \ref{averaged-lvg2} (center). 

\subsubsection{Individual regions \label{individual-excitation}}

In the following we compare the excitation conditions at individual positions.
To do so, we assume that the abundance and velocity gradient
variations do not change with position and fix the CO abundances per
velocity gradient and the fractional
$^{13}$CO abundance to $1\cdot10^{-5}\,{\rm pc}\,(\kms)^{-1}$ and 80 respectively.

\begin{figure}
   \centering
   \includegraphics[width=8.3cm]{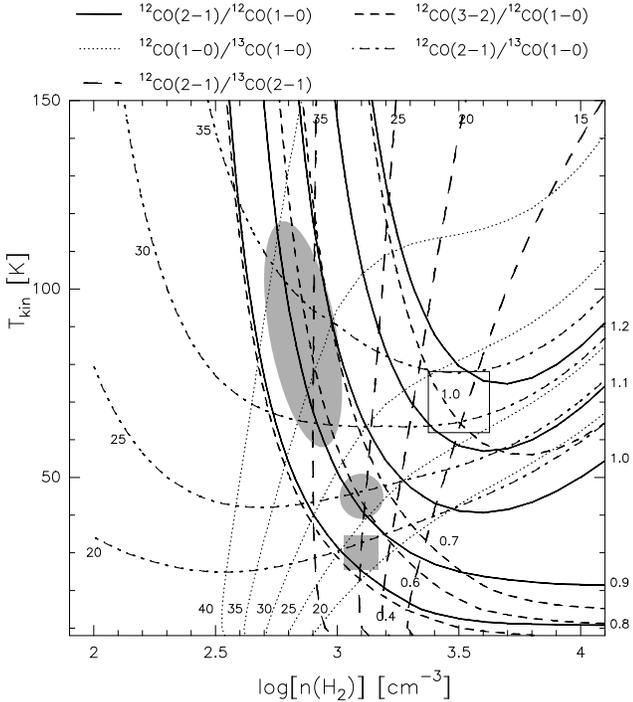}
   \caption{LVG predicted $R_{21}$, $R_{31}$, $R_{10}$, 
   $R^{12/13}_{21}$ and $R^{12/13}_{2 }$ line ratios  as a function of \hh\ density and kinetic
   temperature for a CO abundance per velocity gradient of
   $1\cdot10^{-5}$\,pc\,(\kms)$^{-1}$ and a fractional $^{13}$CO
   abundance of 80. Numbers indicate the respective line ratios. 
   The symbols show the density -- temperature
   combination of the streamers S1 \& S2 (grey circle), the northern
   outflow (grey square), the base of the southern outflow (white
   square) and regions south of the optical disk (S3, S4, and the end
   of the southern outflow; grey ellipse).}
   \label{lvg-region}
\end{figure}

Along the streamers S1 \& S2 the excitation conditions are fairly
similar. The decrease of the $R_{31}$ along S1 is reflected in 
a small decrease of the density and temperature from $log(n_{\hh})=
3.2 {\rm cm}^{-3}$ and $T_{\rm kin}= 50$\,K at the base of the streamer 
to  $log(n_{\hh})= 3.1 {\rm cm}^{-3}$ and $T_{\rm kin}= 40$\,K at its
end. Similar conditions are met along streamer S2 ($log(n_{\hh}) \approx
3.3-3.1 {\rm cm}^{-3}$ and $T_{\rm kin} \approx 45-40$\,K). 
The base of the northern outflow (Pos.\,4)
can be fit well with cooler gas ($T_{\rm kin} \approx 30$\,K) at
comparable density  ($log(n_{\hh})= 3.1\,{\rm cm}^{-3}$) while the high
$R_{10}$ ratio combined with the low \zwco\ ratios, $R_{21}$ and $R_{31}$, at the end of the
outflow leads to $T_{\rm kin} \approx 40$\,K with a low density $log(n_{\hh})= 2.8\,{\rm cm}^{-3}$.
In contrast, the base of the southern outflow consists of denser gas at 
higher temperature ($T_{\rm kin} \approx 70$\,K, $log(n_{\hh}) \approx
3.5\,{\rm cm}^{-3}$), comparable to conditions in the starburst disk
(see Sec. \ref{center}). At the end of the southern outflow as well
as along the streamers S3 and S4 (the positions with the highest
$R_{10}$ ratio) a warm diffuse medium is required to match the
observed line ratios. The temperatures and densities range from 
$T_{\rm kin} \approx 60\,-\,120$\,K and  $log(n_{\hh})= 2.7-3.0\,{\rm
cm}^{-3}$ respectively. Note that the large difference in $R_{21}$ and
$R_{31}$  at the end of the southern outflow (Pos.\,7; 1.1 and 
0.5 respectively) can not be fit by a single gas component. 
In Fig.\,\ref{lvg-region} we show where the different
regions lie in the LVG temperature density space. Not all line ratios at
a given positions can be fit equally well and the locations of the
regions show roughly the best agreement with the measured line ratios. 

\section{Discussion}

\subsection{Properties of the streamer/outflow gas}

The high isotope line ratio observed in the $J=1-0$ transition
($R_{10}$) at all positions clearly rules out that the gas conditions in the outer
parts of M\,82 are similar to those in the disk of the Milky Way or other spiral galaxies.
Typically the \aco\ line has large optical depth in disk clouds in
spiral galaxies and even the emission in \draco\ is only moderately
optically thin which results in $R_{10}$ typically about 5. $R_{10}$ 
much closer to the [CO]/[$^{13}$CO] abundance ratio is a well studied signature
of molecular gas in the centers of active galaxies (including M\,82's
 starburst disk with $R_{10}\approx20$) and mergers (e.g. Aalto \etal\ 
\cite{aalto91}, Sage \& Isbell \cite{sage91}, Henkel \& Mauersberger 
\cite{henkel93}). The reason for the \drco\ depression in luminous
starburst galaxies is still under debate but most explanations involve 
\drco/\zwco\ abundance ratio variations and opacity effects 
(see e.g. Henkel \& Mauersberger \cite{henkel93}, Aalto \etal\ \cite{aalto95}, Kikumoto \etal\ 
\cite{kikumoto98}, Tanigucchi, Ohyama \& Sanders \cite{taniguchhi99},
Paglione \etal\ \cite{paglione01}, Meier \& Turner \cite{meier04}).  
Given its origin from the star forming disk and the likely interaction 
with outflowing ionized gas, high $R_{10}$ ratios are therefore
expected from the gas located in northern and southern molecular
outflow. The observed increase of $R_{10}$ along both outflows
supports the idea that 'superwinds' have large isotope line
ratios ( Tanigucchi, Ohyama \& Sanders \cite{taniguchhi99}). 
For gas along the streamers S1 and S2 the high $R_{10}$ ratios are 
unexpected as $R_{10}$ is usually found to decrease with galactocentric distance (Paglione
\etal\ \cite{paglione01}) and since both streamers are located well outside the
central star forming regions. We note, however, that decreases of
$R_{10}$ typically only becomes significant for galactocentric distances larger
than 2\,kpc (Paglione \etal\ \cite{paglione01}). Increasing $R_{10}$
ratios in the vicinity of starburst have also been observed in e.g. 
IC\,342 (Wright \etal\ \cite{wright93}) and NGC\,6946 (Meier \& Turner \cite{meier04}).
In these cases the large isotopic line ratio has been attribute to 
dispersed low--density gas in the 'spray regions' of barred potentials.
Given the low density in S1 \& S2 and that these streamers
form the base of the disrupted \hi\ features within M82's disk (Yun \etal\ \cite{yun93}) 
we speculate that their large $R_{10}$ may result from dispersed gas due
to tidal forces along S1 \& S2. 

In the context of our LVG models, the large isotopic line ratio is
mainly a result of the very low optical depth of the ground transition of 
$^{13}$CO ($\tau_{13} \approx 0.01-0.05$). These low opacities result
from the low $^{13}$CO abundances per velocity gradient of our models.
Outside the optical disk, in regions with $R_{10} > 30$, higher kinetic temperatures and 
lower \hh\ densities (resulting in lower column densities per velocity
interval) push $^{13}$CO opacities to the lower end of the range.
For gas in local thermodynamic equilibrium (LTE) and large optical 
depth in \aco\ the ratio is given by $R_{10}\approx\tau_{13}^{-1}$.
Although (i) the excitation temperatures between \zwco\ and \drco\ differ
up to a factor of 2 in our models (non LTE) and (ii) the moderate opacities
in \aco\ introduce an additional deviation from the above equation, the
opacity in $^{13}$CO is the most important quantity for
variations in $R_{10}$. This suggests that the CO emission outside 
the optical disk (streamers S3 \& S4 and at the end of both outflows) 
arises from diffuse and presumably warm gas with low opacities in \drco.

Interestingly, the southern and northern outflows have
different excitation conditions (mainly reflected in the low
$R_{21}$ and $R_{31}$ ratios in the northern outflow). The 
lower temperatures towards the north might indicate that the
gas is less affected by outflowing ionized gas. Indeed, the
outflow seen in H$\alpha$ and X-rays is not symmetrical with respect to
the optical disk but brighter and also more extended towards the south
(e.g. Lehnert, Heckman \& Weaver \cite{lehnert99}). This, however,
might simply be an effect of the orientation of the outflow with
respect to the observer since the southern outflow is 
inclined towards us (see e.g. McKeith \etal\ \cite{mckeith95}, 
Greve \cite{greve04} for the outflow geometry). 

Although the kinetic temperatures of the molecular gas are difficult
to constrain without better estimates of the CO abundances, the
general picture to emerge from the LVG models is that most of 
the gas in the outer regions of M\,82 has low densities and low
optical depth at temperatures in excess of 30\,K.

\subsection{Comparison to M\,82's center \label{center}}

The physical conditions of the gas within the central
molecular toroid have been the subject of many studies (e.g. Wild \etal\ \cite{wild92}, G\"usten \etal\
\cite{guesten93}, Mao \etal\ \cite{mao00}, Petitpas \& Wilson
\cite{petit00}, Wei\ss\ \etal\ \cite{weiss01}, Ward \etal\ \cite{ward03}).
These studies have shown, that the observed line ratios can be 
reproduced by emission from a low (LE) and a high excitation (HE) gas component (G\"usten \etal
\cite{guesten93}, Mao \etal\ \cite{mao00}, Wei\ss\ \etal\
\cite{weiss01}, Ward \etal\ \cite{ward03}). In this picture the high excitation
component, responsible for the excitation of levels beyond the
$J=4\to3$ transition, arise from dense gas at temperatures similar to or 
higher than the dust temperature ($log(n_{\hh})\approx 3.5-4.5\,{\rm cm}^{-3}$;
$\tkin \ge 40\,K$). The low excitation component is emitted by diffuse
gas at much lower density ($log(n_{\hh})\approx 3.0 \,{\rm
cm}^{-3}$). 

Our analysis shows that the gas surrounding M\,82's starburst region
has similar properties as the low excitation component in the
starburst center itself. Similar conclusions have been reached from an analysis
of the \zwco\ $J$=3$\to$2/$J$=2$\to$1, \drco/\zwco\  $J$=3$\to$2 and
C$^{18}$O/\zwco\ ratios in the molecular ``halo'' 
by Seaquist \& Clark
(\cite{seaquist01}). They interpret their decreasing line ratios as decreasing CO
excitation and optical depth with increasing distance from the
nuclear region. These findings
are in line with our results. Seaquist \& Clark find that the
variations of the observed line ratios in the transition region between the
center and the ``halo'' are consistent with a decreasing filling
factor of the high excitation component with distance from the
center. In this picture the CO emission from the outer regions arise
exclusively from gas with similar properties than the low excitation 
component in the center. We discuss this scenario in more detail in the next section.
This suggests that mainly the diffuse low density molecular gas in the
starburst region is involved in the outflow. Given the higher inertia of dense (and
therefore more massive) gas clouds, we speculate that the superwind affects the
kinematics of gas at low density more strongly. However, the high CO excitation found at the base of the
southern outflow indicates that also gas at higher density might
be pushed out of the starburst regions, at least to scale heights of a
few hundred pc, less than half the distance seen in the outflow of the
diffuse molecular gas.

\subsection{CO-Line-SED \label{cosed}}

\subsubsection{High $J$ CO fluxes}

We have used published maps in various \zwco\ transitions towards
M\,82 to estimate the total flux density emitted in each line from the center
and the outer parts of the CO distribution. Large--scale maps
(covering the streamer/outflow system) only exist up to the \cco\
transition. The observed flux densities from the outer parts of M\,82 were determined by 
masking the central molecular disk in the \aco\ and \bco\ IRAM 30m data
cubes from Wei\ss\ \etal\ (\cite{weiss01}) and the \cco\ data cube from
Dumke \etal\ (\cite{dumke01}) (after smoothing all data cubes to 22$''$
spatial resolution and regridding onto the same grid). The mask was 
chosen to contain a flux density of $\approx5000$ Jy\,\kms\ in the \aco\
data cube (see Fig.\ref{positions}), close to the value determined by
us for the central molecular disk (Walter, Wei\ss\ \& Scoville \cite{walter02}). CO flux
densities from the center were determined within the inverted mask. For the 
center we also analyzed the \dco\ and \gco\ cubes published by
Mao \etal\ (\cite{mao00}). Since the \dco\ map is larger than our mask
we applied the same technique as for the lower $J$ lines to determine
the central flux density in \dco. The \gco\ map is smaller than the central 
mask and we adopt here the total flux density in the data cube to be
representative for the flux density from the center. The central flux
density of the \fco\ transition was taken from the map by Ward \etal\ (\cite{ward03}),
which roughly covers the central mask.  
The observed \zwco\ flux densities for the center and the outer parts are 
summarized in Table\,\ref{cofluxes}.

\begin{table*}
\caption[]{Observed and model predicted \zwco\ integrated flux densities in the center and outer parts of M\,82.}
\label{cofluxes}
\begin{tabular}{r c c r r r c c}
\hline
\noalign{\smallskip}
 transition& center & outer &model center&model outer &model total&$I_{center}/I_{total}$ & Ref.\\ 
& [Jy\,\kms] & [Jy\,\kms] & [Jy\,\kms] & [Jy\,\kms]& [Jy\,\kms]& [\%] & \\
\noalign{\smallskip} \hline
CO(1--0) & 5.1E3\,$\pm$\,5.0E2 &  1.5E4\,$\pm$\,2.8E3 & 6320 & 17100&23420&27 &1\\
CO(2--1) & 2.2E4\,$\pm$\,5.0E3 &  7.0E4\,$\pm$\,1.8E4 & 24300& 65710&90010&27 &1\\
CO(3--2) & 4.9E4\,$\pm$\,8.0E3 &  9.7E4\,$\pm$\,2.0E4 & 46090&103040&149130&31 &2\\
CO(4--3) & 6.4E4\,$\pm$\,1.3E4 &         --           & 64210& 90570&154780&41 &3\\
CO(5--4) & --                  &         --           & 78110& 45310&123420&63 & \\
CO(6--5) & $^a$\,9.7E4\,$\pm$\,2.4E4 &         --     & 87360& 13650&101010&86 &4\\
CO(7--6) & $^a$\,7.4E4\,$\pm$\,2.3E4 &         --     & 78920&  2960&81880&96 &3\\
CO(8--7) & --                  &         --           & 45570&470&46040&99 &\\
CO(9--8) & --                  &         --           & 13210&    43&13250&100 & \\
\end{tabular}
\begin{list}{}{}
\item[] Ref. 1: Wei\ss\ \etal\ \cite{weiss01} (30\,m data), 2: Dumke \etal\
\cite{dumke01}, 3: Mao \etal\ \cite{mao00}, 4: Ward \etal\ \cite{ward03}
\item[$^{\mathrm{a}}$] total flux covered by the observations\\
\end{list}
\end{table*}

\subsubsection{CO line SED of the starburst disk \label{cosed-center}}
In Fig.\,\ref{co-sed} (left) we show the observed CO line SED (CO
integrated flux density vs. the rotational quantum number) in {\em the
center} of M\,82. CO flux densities rise up to the $J$=6$\to$5
transition, although we can not rule out that
the true \gco\ flux density is higher due to
the limited coverage of the $J$=7$\to$6 map. In this plot we 
also show the LVG predicted CO line SED for a high (HE) and a low
excitation (LE) component in the center. Following the approach by
Seaquist \& Clark we used the LVG model derived from the average 
line ratios in the steamer/outflow system as LE component (LE:
$log(n\,_{\hh})= 3.2\,{\rm cm}^{-3}$, $T_{\rm kin} = 55$\,K, CO
abundance per velocity gradient $1\cdot10^{-5}$\,pc\,(\kms)$^{-1}$; see Sec.\,\ref{average-excitation}). 
Adding a HE component with $log(n\,_{\hh})= 4.2\,{\rm cm}^{-3}$,
$T_{\rm kin} = 50$\,K, similar to conditions in the molecular
lobes (e.g. Wei\ss\ \etal\ \cite{weiss01}), results in a good fit to the
observed CO line SED of the center. For these models we adopted a filling factor of 1/20 and 1/15 for the HE
and the LE component in a solid angel of $\Omega_{\rm{center}}\approx
1000\ \rm{arcsec}^2$ (as defined by our mask for the center). 
For comparison we show the CO line SED in the 
center of M\,82 predicted by a high resolution
LVG analysis based on the $J=1-0$ and $J=2-1$ transitions only (Wei\ss\ \etal\ \cite{weiss01}). 
The fluxes have been calculated taking all individual LVG solutions
within the center into account. This example shows, that LVG models
based on the low--$J$ lines of CO and their rare isotopes give
surprisingly good estimates of the flux density in the high--$J$ \zwco\ lines.

\subsubsection{CO line SED of total emission}
\begin{figure*}
   \centering
   \includegraphics[width=18.3cm]{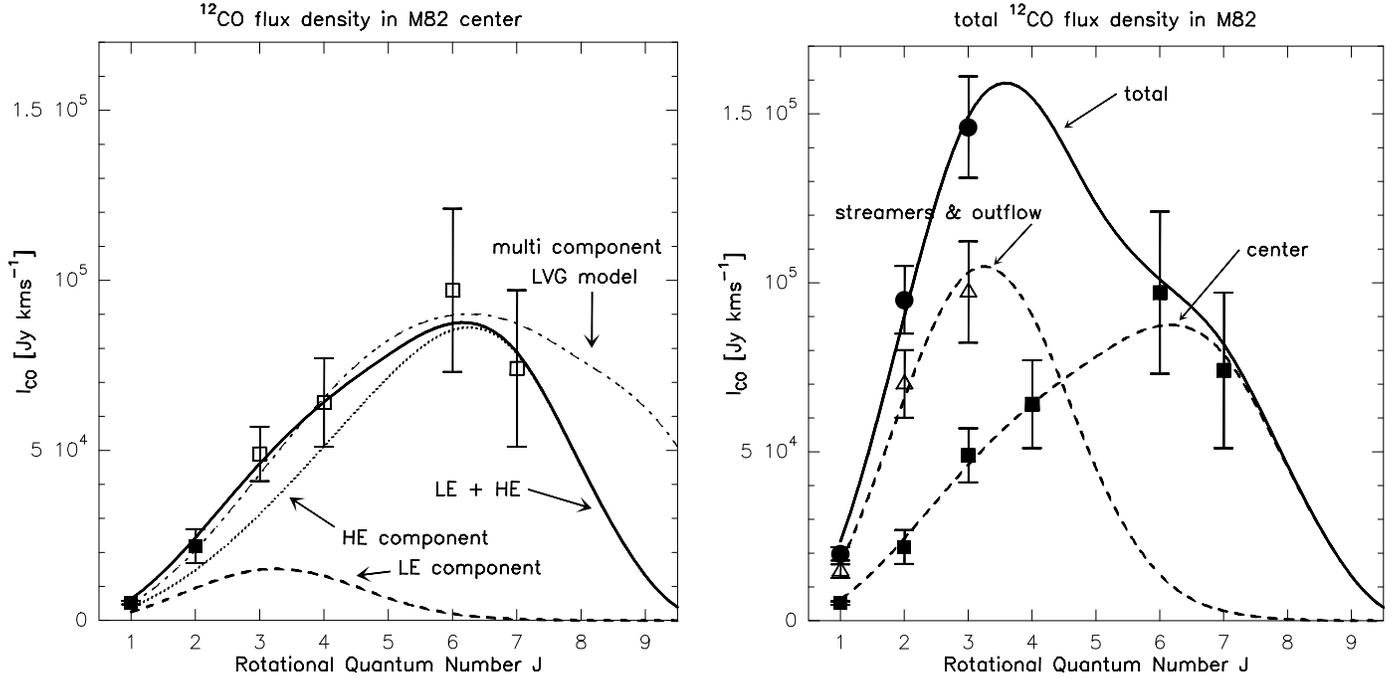}
   \caption{{\it Left:} Observed and model predicted \zwco\ line SED
   for the center of M\,82 using a two component LVG model. 
   The dotted line shows the predicted CO line SEDs of a high 
   (HE), the lower dashed line of a low (LE) excitation component. 
   The LE model correspond to the physical conditions determined in
   the outer regions of the CO distribution, the HE component is
   similar to the conditions in the molecular lobes. The black line is 
   the sum of both components. 
   For comparison we also show the CO line SED for a multi component
   LVG model (dashed line, Wei\ss\ \etal\ \cite{weiss01}).
   The filled squares are those \zwco\ observations which have been used to
   constrain the multi component model. Open squares are CO flux densities
   compiled from the literature. 
  {\it Right:} CO line SED for M\,82 including the molecular streamers and
   outflow. The dashed lines show the model for the center (same as solid
   line to the left) and for the streamers and
   outflow. The solid line 
  shows the total predicted CO line SED in the central $3\times3$ kpc. 
 Filled rectangles are observed flux densities from
   the center of M\,82, open circles the observed flux densities in the
   streamer/outflow system and filled circles the total observed flux density
   in the central $3\times3$ kpc of M\,82. Observed and model predicted
   line flux densities are given in Table\,\ref{cofluxes}.}
   \label{co-sed}
\end{figure*}

In a similar way we have calculated the CO line SED for the
streamer/outflow system. Here we used the LE component with an area 
filling factor of 1/30 (derived from the average \aco\ spectrum at
22$''$ resolution and the LVG predicted line brightness temperature) 
and a source solid angle of $\Omega_{\rm{outer}}\approx
14000\,\rm{arcsec}^2$ 
(estimated from the zero--spacing corrected \aco\ OVRO map). The predicted line flux densities
are shown together with the observed values in Fig.\,\ref{co-sed} (right). 
The plot demonstrates that this simple two phase model for the
center and the outer regions is in good agreement with the observations. 
More detailed studies of the gas in the center show that the
LE component is most likely warmer than the 55\,K used
here (e.g. Wei\ss\ \etal\ \cite{weiss01}, Ward \etal\
\cite{ward03}). Higher temperatures at sightly lower density, however, do not 
change the predicted flux densities in the LE component significantly.
In any case the line SED in the center is dominated by the dense gas
that excites the high--$J$ CO lines. 
This is in line with CO models for the center
of NGC\,253, where the CO line SED is fit by a single high
excitation gas component (Bradford \etal\ \cite{bradford03}, Bayet
\etal\ \cite{bayet04}). 

The {\em combined CO line SED} from the center and the streamer/outflow 
system, which would be observed if M\,82 was shifted to cosmological distances,
however differs strongly from that derived from the center alone (see
Fig.\,\ref{co-sed} right). 
The large area of the surrounding gas makes it the dominant component 
in the emission of the low--$J$ CO transitions. Since this gas is
not dense enough to produce significant flux beyond the \eco\ line,
its CO line SED peaks already at the
\cco\ transition. Consequently, even though two-thirds of the
\aco\ flux density arises from regions outside the star forming disk, CO lines above \eco\ transition
are mainly emitted from the center. In the combined CO line SED, flux densities rise up to the
$J$=4$\to$3 transition and decrease beyond this line. We summarize the CO
flux densities derived from the LVG models in Table \ref{cofluxes}. Table
\ref{cofluxes} also gives the fraction of the total flux density emitted by the center
in each CO transition. For comparison with other studies we give the
model predicted line fluxes in Table \ref{cointensities}.

\begin{table}
\caption[]{LVG model predicted \zwco\ fluxes in the center
($\Omega=2.35\cdot 10^{-8}$\,sr) and outer
($\Omega=3.29\cdot10^{-7}$\,sr) parts of M\,82.}
\label{cointensities}
\begin{tabular}{c c c c}
\hline
\noalign{\smallskip}
 transition&center&outer &total \\ 
 & \multicolumn{3}{c}{10$^{-16}$\,W m$^{-2}$} \\
\noalign{\smallskip} \hline
CO(1--0)  & 0.2& 0.7 &0.9\\
CO(2--1)  & 1.9& 5.1 &6.9\\
CO(3--2)  & 5.3&11.9 &17.2\\
CO(4--3)  & 9.8&13.9 &23.8\\
CO(5--4)  &15.0& 8.7 &23.7\\
CO(6--5)  &20.2& 3.1 &23.3\\
CO(7--6)  &21.2& 0.8 &22.0\\
CO(8--7)  &14.0& 0.1 &14.2\\
CO(9--8)  & 4.6& 0.01& 4.6\\
\end{tabular}
\end{table}

\subsection{CO line SEDs at high redshift}

Currently CO line SEDs in sources at redshift $>\,2$ are still
poorly constrained. A few sources, however, have been observed in 
more than two CO transitions which allow at least for a quantitative
comparison with the CO line SED of M\,82. Among the best studied
examples are the Cloverleaf quasar ($z=2.5$, Barvainis \etal\
\cite{barvainis97}, Wei\ss\ \etal\ \cite{weiss03}), IRAS FSC\,10214
($z=2.3$, Brown \& Vanden Bout \cite{brown92}, Solomon, Downes \& Radford
\cite{solomon92}), PSS\,2322+1944 ($z=4.1$, Cox \etal\ \cite{cox02}, Carilli \etal
\cite{carilli02a}), BR1202-0725 ($z=4.7$ Omont \etal\ \cite{omont96},
Carilli \etal\ \cite{carilli02b}) and J1148+5251 ($z=6.4$, Walter
\etal\ \cite{walter03}, Bertoldi \etal\ \cite{bertoldi03}).
In all these sources the CO flux densities are rising with increasing rotational quantum
numbers (at least up to the $J=6\to5$ transitions), similar to the center of
M\,82 and NGC\,253. These studies 
argue against large extended low excitation CO halos surrounding these
objects at high redshifts. However, there are a few examples where the
flux density from the CO $J$=3$\to$2 transition is similar or higher than that of the $J$=7$\to$6 line (see
e.g. SMM\,14011+0252 ($z=2.5$, Downes \& Solomon \cite{downes03}),
SMM\,16368+4057 ($z=2.4$, Neri \etal\ \cite{neri03}). In light of our
new results in M\,82 we suggest that the low $J$=7$\to$6/$J$=3$\to$2 ratios
in these sources may not necessarily be caused by a different excitation of 
the central molecular gas concentration, but may result from an
additional more extended and diffuse gas reservoir.

\section{Summary}

We have presented observations of the $J$=1$\to$0 and $J$=2$\to$1 transition
of \zwco\ and \drco\ towards 13 selected regions outside the central
starburst disk of M\,82 covering the prominent molecular streamers and
the CO outflow. Our observations show that the $J$=2$\to$1/$J$=1$\to$0 and
$J$=3$\to$2/$J$=1$\to$0 ratios of \zwco\ are lower in all streamer/outflow regions
than in the central molecular disk. The \zwco\ ratios are fairly
constant along the analyzed positions and show only a slight trend of
decreasing ratios with increasing distance from the center. The
isotope line ratio \aco/\draco\ is around 25 along the most prominent
molecular streamers, similar to values found in the starburst disk,
and increases along the outflows to 35. Similarly high values for this
ratio are found in the diffuse streamers S3 and S4 south-east of the
optical disk of M\,82.

An LVG analysis of the gas suggests that the low \zwco\ ratios
combined with the high isotope line ratio are best explained by a low
density ($log(n_{\hh}) \approx 3.0\,{\rm cm}^{-3}$) molecular
environment with kinetic temperatures in excess of 30\,K and
[\drco]/[\zwco] abundance ratios $\sim 80$. In this picture the high
\aco/\draco\ ratio is mainly caused by very low opacities in the \drco\
lines. Our analysis does not allow to rule out denser gas at lower
kinetic temperature if abundance variations are taken into
account. However densities in excess of $log(n_{\hh}) \approx
3.5\,{\rm cm}^{-3}$ require [\zwco]/[\drco] abundance ratios $>100$ or
very small \zwco\ abundances per velocity gradient ([CO]/\gradv$ <
1\cdot10^{-6}$) to explain the observed line ratios which we consider
unlikely.

We use the LVG predicted line brightness temperature ratios of the
\zwco\ $J>$3 transitions to estimate the total flux density emitted in the
high--$J$ lines from the diffuse CO component surrounding M\,82
starburst region. We find that the density of the streamer/outflow gas is
not high enough to produce significant flux above the $J$=5$\to$4
line. As a consequence the CO-line SED of the outer gas drops off
beyond $J$=3$\to$2 line while it is rising at least up to the
$J$=6$\to$5 line in the central molecular disk.  The total $J$=1$\to$0 and
$J$=2$\to$1 flux density within the central $3\time3$kpc, however, is clearly
dominated by the emission from the outer regions.  This implies that
one has to exercise caution when comparing global, unresolved, high--z
line ratios to the values measured in the {\em centers only} of nearby
starburst galaxies. We conclude that small high--$J$/low--$J$ CO ratios
ratios seen in some high--z sources are not necessarily caused by a different
excitation of the central molecular gas concentration, but may result
from an additional more extended and diffuse gas reservoir around these
systems.
 
\begin{acknowledgements}
We thank the referee, Dr. M.\,S. Yun, for valuable suggestions which
helped to improve this paper. We also thank Drs. M. Dumke and R.\,Q. Mao for 
making available their data to us. This paper is based on observations
carried out at the IRAM 30\,m telescope.
IRAM is supported by INSU/CNRS (France), MPG (Germany) and IGN (Spain).
     
\end{acknowledgements}


\begin{thebibliography}{}
\bibitem[1991]{aalto91} Aalto, S., Black, J.H., Johansson, L.E.B., \& Booth, R.S. 1991, A\&A, 249 323
\bibitem[1995]{aalto95} Aalto, S., Booth, R.S., Black, J.H., \& Johansson, L.E.B. 1995, A\&A, 300, 369
\bibitem[1997]{barvainis97} Barvainis, R., Maloney, P., Antonucci, R., \& Alloin, D. 1997, ApJ, 484, 695
\bibitem[2004]{bayet04} Bayet, E., Gerin, M., Phillips, T.G., \& Contursi, A. 2004, A\&A, 427, 45
\bibitem[2003]{bertoldi03} Bertoldi, F., Cox, P., Neri, R., \etal\ 2003, A\&A, 409, L47
\bibitem[2003]{bradford03} Bradford, C.M., Nikola, T., Stacey, C.J.,
Bolatto, A.D., Jackson, J.M., Savage, M.L., Davidson, J.A., \&
Highdon, S.J. 2003, ApJ, 586, 891  
\bibitem[1992]{brown92} Brown, R.L, \& Vanden Bout, P.A. 1992, ApJ, 397, L19
\bibitem[2002a]{carilli02a} Carilli, C.L., Kohno K., Kawabe, R., \etal\ 2002a, AJ, 123, 1838
\bibitem[2002b]{carilli02b} Carilli, C.L., Cox, P., Bertoldi, F., \etal\, 2002b, ApJ, 575, 145
\bibitem[1997]{cheng97} Cheng, K.-P., Collins, N., Angione, R., Talbert, F., Hintzen, P.,
Smith, E.P., Stecher, T., and the UIT Team, 1997, `UV/Visible Sky
Gallery'
\bibitem[2002]{cox02} Cox P., Omont, A., Djorgovski S.G., \etal\ 2002, A\&A, 387, 406
\bibitem[2003]{downes03} Downes, D., \& Solomon, P.M. 2003, ApJ, 528, 37
\bibitem[2001]{dumke01} Dumke, M.,  Nieten, C., Thuma, G., Wielebinski, R. \&  Walsh, W.\ 2001, A\&A, 373, 853
\bibitem[1993]{guesten93} G\"usten, R., Serabyn, E., Kasemann, C.,
Schinckel, A., Schneider, G., Schulz, A., \& Young, K. 1993, ApJ, 402, 537
\bibitem[1998]{greve98} Greve, A., Kramer, C., \& Wild, W.\ 1998, A\&A, 133, 271
\bibitem[1998]{greve04} Greve, A. 2004, A\&A, 416, 67
\bibitem[1993]{henkel93} Henkel, C., \& Mauersberger, R. 1993, A\&A, 274, 730
\bibitem[1994]{johansson94} Johansson, L.E.B, Olofsson, H.,
Hjalmarson, A., Gredel, R., \& Black, J.H. 1994, A\&A, 291, 89
\bibitem[1998]{kikumoto98} Kikumoto, T., Taniguchi, Y., Nakai, N.,
Ishizuki, S., Matsushita, S., \& Kawabe, R.\ 1998 PASJ, 50, 309 
\bibitem[1999]{lehnert99} Lehnert, M.D., Heckman, T.M., \&  Weaver,
K.A. 1999, ApJ, 523, 575 
\bibitem[2000]{mao00} Mao, R.Q., Henkel, C., Schulz, A., \etal\ 2000, A\&A, 358, 433
\bibitem[1995]{mckeith95} McKeith, C.D., Greve, A., Downes, D., \&
Prada, F. 1995, A\&A, 293, 703
\bibitem[2004]{meier04} Meier, D.S., \& Turner, J.L 2004, AJ, 127, 2069
\bibitem[2003]{neri03} Neri, R., Genzel, R., Ivison, R.J., \etal\
2003, ApJ, 597, L113 
\bibitem[1996]{omont96} Omont, A., McMahon, R.G., Cox, P., \etal\, 1996, A\&A, 315, 1
\bibitem[2001]{paglione01} Paglione, T.A.D, Wall, W.F., Young, J.S., \etal\ 2001, ApJSS, 135, 183  
  Diamond, P., Wills, K.A., Wilkinson, P.N., \& Alef, W. 1999, MNRAS, 307, 761
\bibitem[2000]{petit00} Petitpas, G.R., \& Wilson, C.D. 2000, ApJ, 538, L117
\bibitem[1991]{sage91} Sage, L.J., \& Isbell, D.W. 1991, A\&A, 247, 320
\bibitem[2001]{seaquist01} Seaquist, E.R., \& Clark, J. 2001, ApJ, 552, 133
\bibitem[1992]{solomon92} Solomon, P.M., Downes, D. \& Radford. S.J.E. 1992, ApJ, 398, L29  
\bibitem[1984]{stark84} Stark, A.A., \& Carlson, E.R. 1984, ApJ, 279, 122
\bibitem[1999]{taniguchhi99} Tanigucchi, Y., Ohyama, Y., \& Sanders, D.B. 1999, ApJ, 522, 214
\bibitem[2001]{taylor01} Taylor, C.L., Walter, F., \& Yun, M.S. 2001, ApJ, 562, L43
\bibitem[2003]{walter03} Walter, F., Bertoldi, F., Carilli, C.L. \etal\, 2003, Nature, 424, 406
\bibitem[2002]{walter02} Walter, F., Wei\ss, A., \& Scoville, N.Z. 2002, ApJ, 580, L21
\bibitem[2003]{ward03} Ward, J.S., Zmuidzinas, J., Harris, A. \& Isaak, K.G. 2003, ApJ, 587, 171
\bibitem[1999]{weiss99} Wei\ss, A., Walter, F., Neininger, N., \& Klein, U. 1999, A\&A, 345, L23
\bibitem[2001]{weiss01} Wei\ss, A., Neininger, N., H\"uttemeister, S., \& Klein, U. 2001, A\&A, 365, 571
\bibitem[2003]{weiss03} Wei\ss, A., Henkel, C., Downes, D., \& Walter, F. 2003, A\&A, 409, L41
\bibitem[1992]{wild92} Wild, W., Harris, A.I., Eckart, A., \etal\ 1992, A\&A, 265, 447
\bibitem[1993]{wright93} Wright, M.C.H., Ishizuki, S., Turner, J.L., Ho, P.T.P., \& Lo, K.Y. 1993, ApJ, 406, 470
\bibitem[1984]{young84} Young, J.S., \& Scoville, N.Z. 1984, ApJ, 287, 153
\bibitem[1993]{yun93} Yun, M.S, Ho, P.T.P., Lo, K.Y. 1993, ApJ, 411, 17
\bibitem[2000]{yun00} Yun, M.S, Carilli, C.L., Kawabe, R. \etal\, 2000, ApJ, 528, 171

\end{thebibliography}
\end{document}